\documentstyle[12pt]{article}
\begin{document}
\title{{\bf Evidence Against Astrophysical Dyadospheres}
\thanks{Alberta-Thy-06-06, astro-ph/0610340,
accepted 2006 August 17 by The Astrophysical Journal}}
\author{
Don N. Page
\thanks{Internet address:
don@phys.ualberta.ca}
\\
Institute for Theoretical Physics\\
Department of Physics, University of Alberta\\
Room 238 CEB, 11322 -- 89 Avenue\\
Edmonton, Alberta, Canada T6G 2G7\\
and \\
Asia Pacific Center for Theoretical Physics (APCTP)\\
Hogil Kim Memorial Building \#519\\
POSTECH, San 31\\
Hyoja-dong, Namgu, Pohang\\
Gyeongbuk 790-784, Korea
}
\date{(2006 August 15, minor changes for the eprint added Oct. 11)}

\maketitle
\large
\begin{abstract}
\baselineskip 14.5 pt

It is shown how pair production itself would almost certainly prevent the
astrophysical formation of macroscopic dyadospheres, hypothetical regions,
extending many electron Compton wavelengths in all directions, where the
electric field exceeds the critical value for microscopically rapid Schwinger
pair production.  Pair production is a self-regulating process that would
discharge a growing electric field, in the example of a hypothetical collapsing
charged stellar core, before it reached 6\% of the minimum dyadosphere value,
keeping the pair production rate more than 26 orders of magnitude below the
dyadosphere value, and keeping the efficiency below $2\times
10^{-4}\sqrt{M/M_\odot}$.

\end{abstract}
\normalsize

\section{Introduction}

Ruffini and his collaborators
\cite{DR,R1, PRX1,RSWX1,RX,R2,R3,RSWX2,RSWX3,R4,R5,R6,RSWX4,PRX2,R7,PRX3,BRX1,
RBFCX,RBFXC,RV,RXBFC,RBCFX1,R8,R9,R10,R11,CMRX,RBCFX2,RBBCFX1,RVX1,RVX2,RVX3,
RVX4,RBCFVX,RBCFX3,RBBCFX2,RBCFGX1,RFVX1,RBCFGX2,BBCFRX1,BR1,CBBCFRX,FBBCRX,
BR2,BR3,RFVX2,RBBVXCFG,RBBCFGLVX,RBBCFGVX,BBCFRX2,XRFG,RBBCFX3,RBBCFGX,Vag}
have proposed a model for explaining gamma ray bursts that presumes the initial
existence of what they call a {\it dyadosphere}, a macroscopic region of
spacetime where the electric field exceeds the critical electric field $E_c
\equiv m_e^2 c^3/(\hbar e) \approx 1.32 \times 10^{16} \:\mathrm{V/cm}$ for
microscopically rapid Schwinger pair production \cite{Sch1,Sch2,Sch3,Sch4,Nik}. 
The difficulty of producing these large electric fields is a problem with this
model that has not been adequately addressed.

There are at least two strong reasons for doubting that such large electric
fields can develop over astrophysical scales in all directions (i.e., over
length scales much larger than the electron Compton wavelength, or much larger
than the collision regions of individual charged particles).  First, it would
be very difficult to develop sufficient charge imbalance for macroscopic
electric fields to produce significant numbers of pairs.  Second, even if
macroscopic pair production could somehow be achieved, I shall show in this
paper that this process is sufficiently self-regulating that it prevents the
electric field from achieving a value that would produce pairs at even
$10^{-26}$ that of dyadosphere models, assuming that the field is extended over
a three-dimensional region at least as large as the Schwarzschild radius of a
solar-mass black hole and that the field develops over a time that is at least
as long as the corresponding timescale.  I conclude that it is highly
implausible that macroscopic dyadospheres can form in outer space, and
therefore invoking them for models of gamma ray bursts is not at all likely to
be viable.

The first reason for being extremely doubtful of the existence of astrophysical
dyadospheres is that it is very difficult for a large charge imbalance to
develop over a macroscopic region astrophysically, because of the very high
charge-to-mass ratio of elementary particles.  For example, the ratio of the
electrostatic repulsion to the gravitational attraction of two protons, each of
charge $q$ and mass $m_p$, is the square of their charge-to-mass ratio in
Planck units, which is
 \begin{equation}
 \left({q\over m_p}\right)^2 \equiv {q^2\over 4\pi\epsilon_0 G m_p^2}
 \approx 1.24\times 10^{36}.
 \label{eq:a}
 \end{equation}
This implies that if one had a spherical object, such as a stellar core, with a
positive charge-to-mass ratio $Q/M$ greater than the inverse of the
charge-to-mass ratio of the proton, $m_p/q \approx 9\times 10^{-19}$, the
electrostatic repulsive force on the protons at the surface would be greater
than the gravitational attractive force, so such protons would most likely be
ejected.  (If the object had a negative charge, electrons of mass $m=m_e$ would
be expelled if $-Q/M > m_e/q \approx 4.9\times 10^{-22}$, lower by the factor
of the mass ratio of the proton and the electron, $m_p/m_e \approx 1836$, so
the maximum value of the charge of such an object would be even less if it were
negative.)

If one takes the mass-to-charge ratio of the proton to be a rough estimate of
the maximum charge-to-mass ratio of an astrophysical object (or else gravity
would not be strong enough to hold in the protons that make up the excess
charge), then using the formulas of the succeeding sections, one can readily
calculate that at the surface of a spherical object of radius $R$ and mass $M$,
the ratio of the electric field value, $E$, to the critical field value of a
dyadosphere, $E_c \equiv m_e^2 c^3/(\hbar q) \approx 1.32 \times 10^{16}
\:\mathrm{V/cm}$, is
 \begin{equation}
 {E\over E_c} \leq {\hbar c \, m_p\over 4 G M_\odot m_e^2}
 \left({M_\odot\over M}\right)\left({2GM\over c^2 R}\right)^2
 < 1.2\times 10^{-13}\left({M_\odot\over M}\right).
 \label{eq:b}
 \end{equation}
Therefore, if protons can be ejected from an astrophysical object whenever the
electrostatic repulsion exceeds the gravitational attraction, then the electric
field is constrained to be more than 13 orders of magnitude smaller than the
critical value for a dyadosphere [if the mass is greater than 1.2 solar masses,
which would be a conservative lower limit on any mass that could contract to
$2GM/(c^2 R) \sim 1$].  For a negatively charged object, the corresponding
limit would be more than 16 orders of magnitude smaller than the critical value
for a dyadosphere.

Although it seems very unlikely to work, one might seek to evade the
electrostatic expulsion of protons by postulating that they are bound by
nuclear forces to an astrophysical object, such as a collapsing neutron star
core.  The critical electric field $E_c$ that Ruffini and his collaborators
used to define the minimum value for a dyadosphere would give the electrostatic
force on an electron or proton of magnitude $F_c = qE_c = (m_ec^2)^2/(\hbar c)
\approx 0.00132$ MeV/fm, whereas nuclear energies of the order of a few MeV
over length scales of the order of a fermi would give nuclear forces of the
order of a few MeV/fm, about 3 orders of magnitude larger than the
electrostatic force of a minimal dyadosphere on a proton.  Ruffini has proposed
\cite{R12} that since the gravitational force is mainly provided by the
nucleons of a collapsing neutron star core, with a tiny fraction of the
nucleons in the form of protons that are bound by nuclear forces to the surface
of the neutron star core, a sufficient fraction can give the electric field
originating the dyadosphere.

My suspicion is that such a configuration would be highly unstable to pieces of
the charged surface breaking off and being ejected by the huge electrostatic
forces on them.  However, I do not have a firm result of when this would
definitely happen, and therefore I was led to do the calculations below with
the second mechanism (self-regulation) for preventing the occurrence of a
dyadosphere.  This, as we shall see below, will almost certainly discharge any
growing electric field well before it reaches dyadosphere values.  This occurs
essentially because astrophysical length scales are much greater than the
electron Compton wavelength, which is the scale at which the pair production
becomes significant at the critical electric field value for a dyadosphere. 
Therefore, the electric field will discharge astrophysically even when the pair
production rate is very low on the scale of the electron Compton wavelength.

In particular, I shall assume that a macroscopic astrophysical electric field
cannot develop faster than the fastest timescale for bulk motions, which I
shall assume to be the final gravitational collapse timescale for the smallest
stellar mass that can collapse into a black hole, of the order of $M_\odot
\approx 5$ microseconds or greater.  If a macroscopic electric field cannot
develop much faster than this by astrophysical processes, the discharge from
the pair production will then self-regulate the field to keep it from getting
near dyadosphere values.

The calculations below lead to the conclusion that it would be very difficult
astrophysically to achieve, over a macroscopic region comparable to the size of
a black hole or larger star, electric field values greater than a few percent
of the minimum value for a dyadosphere, if that.  The Schwinger pair production
itself would then never exceed $10^{-26}$ times the minimum dyadosphere value.

Since the idealized model below does give pair production at macroscopically
significant rates (though more than 26 orders of magnitude below that of a
dyadosphere by the original definition given above), one might revise the
definition of a dyadosphere to include any macroscopic electric field which
gives macroscopically significant pair production.  Then (assuming that
sufficient charge separation can somehow be achieved by forces necessarily much
stronger than gravitational forces, to evade the limitations discussed above),
the arguments and model below would not exclude the possibility of such a
revised concept of a dyadosphere.  However, we shall see below that the much
weaker amount of pair production in the model below does not seem to have
nearly high enough energy efficiency for models of gamma ray bursts.  Therefore
in this paper, unless otherwise stated, I shall stick with the original
definition of a dyadosphere, given above and in more detail in the following
section.

\section{Pair production rate and definition of a dyadosphere}

The Schwinger pair production \cite{Sch1,Sch2,Sch3,Sch4} gives a rate
${\mathcal N}$ of electron-positron pairs per four-volume (per three-volume and
per time) in a uniform electric field that is \cite{Nik}
 \begin{equation}
 {\mathcal N} = {q^2 E^2\over 4\pi^3}\exp{\left(-{\pi m^2\over qE}\right)}
             \equiv {m^4\over 4\pi} {e^{-w}\over w^2},
 \label{eq:7}
 \end{equation}
where (using Planck units $G = \hbar = c = 4\pi\epsilon_0$)  $m \approx
4.185\times 10^{-23}$ is the mass of the electron, $q = \sqrt{\alpha} \approx
0.08542$ is the magnitude of the charge of the electron, and
 \begin{equation}
 w \equiv {\pi m^2\over qE} \equiv {\pi E_c \over E},
 \label{eq:8}
 \end{equation}
if one defines the critical electric field strength $E_c$ to be
 \begin{equation}
 E_c \equiv {m^2 \over q} \approx 1.323 \times 10^{16} \:\mathrm{V/cm}.
 \label{eq:9}
 \end{equation}

Ruffini \cite{R1,PRX1,RSWX1,RX,R2} has defined a {\it dyadosphere} to be a
region of spacetime in which the electric field $E$ is greater than the
critical electric field $E_c$.  Therefore, it is a region in which $w < w_c =
\pi$, and a region in which the electron-positron pair production rate is
 \begin{eqnarray}
 {\mathcal N} > {\mathcal N}_c = {m^4\over 4\pi} {e^{-\pi}\over \pi^2}
   \approx 10^{-93} \approx 4.7\times 10^{48}\:\mathrm{cm}^{-3}\:\mathrm{s}^{-1}
     \approx 7.4\times 10^{58} M_\odot^{-4}.
 \label{eq:10}
 \end{eqnarray}

The last number means that if one had a dyadosphere over a cube of edge length
$M_\odot \approx 1.5$ km and over a time $M_\odot \approx 5$ microseconds, it
would produce more than $7\times 10^{58}$ pairs, and the positrons would have a
total charge $\sim 10^{20} M_\odot$.

This shows that the minimum dyadosphere pair production rate, though only $4\pi
e^{-\pi}$ in units of the electron Compton wavelength, is utterly enormous at
macroscopic astrophysical scales.  This strongly suggests that dyadospheres
will never form over macroscopic astrophysical scales.  Indeed, the
calculations below confirm this for a simple spherically symmetric model and
show that under extremely conservative assumptions, the pair production rate
from astrophysical gravitational collapse is likely to be always less than
$10^{-26}$ of that of a dyadosphere.  (Under the plausible but not rigorous
arguments of the Introduction that the charge is ejected when the electrostatic
repulsion exceeds the gravitational attraction, the upper limit in a region of
the order of the size of a black hole would be trillions of orders of magnitude
weaker than a dyadosphere, a factor of more than $10^{10^{13}}$, and hence
completely negligible.)

\section{Crude upper limit of pair production rate}

One can make the following crude estimate of the upper limit of the pair
production rate from any macroscopic astrophysical process that has an electric
field $E = \pi E_c/w = \pi m^2/(qw)$ extending along the field direction a
macroscopic distance $L \gg L_{\mathrm{pair}} = 2m/(qE) = 2w/(\pi m)$ (where
$L_{\mathrm{pair}}$ is the minimum distance along this field over which it is
energetically possible to produce an electron-positron pair with total rest
mass $2m$), and lasting for a time scale $T \,\stackrel{>}{\sim}\, L/c$.  We
assume here and henceforth that since we are assuming a macroscopic electric
field that extends over a distance much greater than $L_{\mathrm{pair}}$, we
may take it to be nearly homogeneous over the pair-production distance
$L_{\mathrm{pair}}$, so that the pair production rate is given very accurately
by Eq. (\ref{eq:7}) for a uniform field.

This pair production over a time $T$ and length $L$ along the electric field
direction will lead to a charge density produced, per area perpendicular to the
field, of magnitude $\sigma \sim q {\mathcal N} T L$, which will discharge the
field if it becomes comparable to $E/(4\pi)$.  Therefore, the average pair
production rate is roughly limited by
 \begin{eqnarray}
 {\mathcal N} = {m^4\over 4\pi} {e^{-w}\over w^2}
  \,\stackrel{<}{\sim}\, {E\over 4\pi q T L} = {m^2\over 4 q^2 w T L}.
 \label{eq:11a}
 \end{eqnarray}
      
Now if for the given $T$ and $L$ one defines
 \begin{equation}
 X \equiv {m^2 q^2 T L\over \pi}
  \approx {cT L\over (8\times 10^{-10}\,\mathrm{cm})^2} \gg 1,
 \label{eq:11b}
 \end{equation}
the inequality (\ref{eq:11a}) implies that $w e^w \,\stackrel{>}{\sim}\, X$ or
$w \,\stackrel{>}{\sim}\, \ln{X} - \ln{\ln{X}}$.  This then leads to the ratio
of the actual pair production rate ${\mathcal N}$ to the minimal dyadosphere rate
${\mathcal N}_c$ being
 \begin{equation}
 {{\mathcal N}\over{\mathcal N}_c} \,\stackrel{<}{\sim}\,
 {\pi^2 e^\pi\over X(\ln{X}-\ln{\ln{X}})}
  \sim {(10^{-8} \,\mathrm{cm})^2\over cT L} \ll 1
 \label{eq:11c}
 \end{equation}
for any macroscopic time and length scales, which would give $X \gg \pi^2
e^{\pi}$.

This strongly suggests that no matter what the situation, if the time and
length scales for the electric field obey
 \begin{equation}
  cT \,\stackrel{>}{\sim}\, L
  \gg {\pi^{3/2} e^{\pi/2}\over m q} \sim 10^{-8} \,\mathrm{cm}.
 \label{eq:11d}
 \end{equation}
then dyadospheres of such a macroscopic size will not form.

In the following sections, we shall find a more precise upper bound on the pair
production rate in an example with spherical symmetry.  This example confirms
the limit above and strengthens the evidence that pair production rates in
macroscopic processes must almost certainly be many, many orders of magnitude
below dyadosphere rates.

It must be admitted that the bound above is heuristic and is not a rigorous
mathematical result.  For example, if one just wants a hypothetical example of
how a spatially infinite dyadosphere could develop from initial conditions
without one, consider the following collision of two plane electromagnetic
waves with the electromagnetic field tensor
 \begin{equation}
 \mathbf{F} = E\,[\theta(t-z)\mathbf{d}x\wedge\mathbf{d}(t-z)
                          +\theta(t+z)\mathbf{d}x\wedge\mathbf{d}(t+z)],
 \label{eq:11e}
 \end{equation}
where $\theta$ denotes the Heaviside step function.

For $t < 0$, there are two incoming electromagnetic fields, with one front at
$z=t=-|t|$ moving in the positive $z$-direction and the other front at
$z=-t=|t|$ moving in the negative $z$-direction.  These fields are nonzero and
constant behind their fronts, for $|z| > |t|$, each with electric field $E$ in
the positive $x$-direction, but with the field at $z < -|t|$ that is moving in
the positive $z$-direction having its magnetic field, also of magnitude $E$,
pointing in the positive $y$-direction, and with the field at $z > +|t|$ that
is moving in the negative $z$-direction having its magnetic field of magnitude
$E$ pointing in the negative $y$-direction.  Because the scalar invariants of
the electromagnetic field vanish for $t < 0$, there is no pair production then.

However, for $t > 0$, the two incoming electromagnetic fields overlap for $|z|
< t$ and have their magnetic fields cancel out there.  In the absence of pair
production, there would just be an electric field $2E$ in the $x$-direction in
this region, which is infinitely extended in the $x$- and $y$-directions. 
Therefore, if $2E > E_c$, there would then be a dyadosphere that is infinitely
large (in two directions) for $t > 0$, until the pair production reduced the
field below the dyadosphere value.

This hypothetical example circumvents the heuristic argument for the limit
above first by not having the time scale $T \,\stackrel{>}{\sim}\, L/c$, but
more importantly by having the incoming electromagnetic field (at $t < 0$) have
Fourier components of arbitrarily short wavelength, in order to produce the
sharp jump in the field from zero at $|z| < -t = |t|$ to its nonzero values at
$z < -|t|$ and at $z > |t|$.  If instead the fields of the incoming plane waves
rose sufficiently slowly (on the microscopic scale of the electron Compton
wavelength, though this certainly allows what would be considered a rapid
variation on macroscopic astrophysical scales) to dyadosphere values of the
electric field (though the equal magnetic field magnitudes in the perpendicular
direction would prevent the field from actually being a pair-producing
dyadosphere in each incoming plane wave when they are separate), then when the
waves started overlapping to cancel the magnetic field and give pair
production, the pair production itself would regulate the field from ever
attaining dyadosphere values.  Therefore, unless somehow it is possible for
macroscopic astrophysical electromagnetic fields to develop significant
components of microscopically large wavenumbers (and hence significant changes
in the fields over length scales of the order of an electron Compton
wavelength), it seems very unlikely that macroscopic dyadospheres will ever
develop astrophysically.

\section{Schwinger discharge of a charged collapsing core}

In this section we shall analyze the pair production and discharge of an
electric field produced by the collapse of a hypothetical charged sphere or
stellar core, ignoring the processes discussed in the Introduction that would
most probably cause almost all of the excess charge of the sphere to be
electrostatically ejected.

For a sphere collapsing in finite time, there is only a finite time for the
electric field to be discharged by the Schwinger pair production process, so
the discharge is never complete, but instead leaves a residual electric field
at each moment of the collapse.  We shall calculate an upper limit for this
value and show that it is always more than a factor of 18 less than the minimum
value for a dyadosphere.  Because the pair-production rate is exponentially
damped by the inverse of the electric field value, the pair production rate
itself never exceeds a value that is more than 26 orders of magnitude below
that of a dyadosphere.  (The factor of the order of $10^{26}$, which we shall
derive below, comes mainly from the fine structure constant multiplied by the
square of the ratio of the Schwarzschild radius of a solar mass black hole to
the Compton wavelength of an electron, which is why this factor is so large.)

The maximum electric field is smaller for cores that collapse into larger black
holes (because the discharge time during infall is greater), so for a very
conservative upper limit on the electric field, we shall assume that the black
hole that forms has one solar mass.  Of course, we expect that the minimum mass
of a black hole that forms astrophysically has significantly more than one
solar mass, so the corresponding maximum electric field would be weaker (by a
logarithmic factor such that the maximum pair production rate would go
essentially inversely with the square of the mass of the black hole).  That is,
we are assuming that one solar mass is a very conservative lower limit on the
mass of any black hole that forms astrophysically in the present universe (as
distinct from, say, primordial black holes that might have formed much smaller
in the very early universe; our calculations will not apply to those, but they
will have had by now a very long time to discharge and so would also not be
expected to have significant charge today).

The maximum electric field is also smaller the slower that the core collapses
(giving more time for discharge), so again to get a very conservative upper
limit on the electric field, we shall assume that the core falls in as fast as
is astrophysically possible, which is free fall with zero binding energy.  We
assume that it is not astrophysically possible to have the spherical outer
boundary of a collapsing core moving inward with a velocity so high that it
would have come from an unbound configuration with nonzero inward velocity at
radial infinity, or that at smaller radii it could have non-gravitational
inward accelerations.

Thirdly, the maximum electric field is of course smaller the smaller the
initial charge on the collapsing core.  To get the maximum electric field
possible, we shall assume that the initial charge-to-mass ratio of the
collapsing core is unity, the largest value possible for which the
electrostatic repulsive forces do not overwhelm the total gravitational
attractive forces on the entire core and prevent the core from collapsing
(since we are excluding the possibility that the core is shot in from far away
or is otherwise pushed inward by nongravitational forces, which is not at all
astrophysically plausible).

Now of course if the charge-to-mass ratio of the core really were unity, the
core would not fall until there is some discharge.  But merely to get a
conservative upper limit on the electric field as the core collapses, we shall
for simplicity assume that the initial charge equals the mass but ignore the
electrostatic repulsion, so that the core nevertheless falls in at the rate it
would from purely gravitational free fall from infinity, with no reduction of
the free fall rate by the electrostatic repulsion of the charge.  The
gravitational effects of the electric field outside the core would also reduce
the free fall time, but we shall ignore this effect as well and simply take the
external gravitational field to be given by the vacuum spherically symmetric
Schwarzschild metric.

First we shall make a rough Newtonian estimate of the discharge rate for a
collapsing charged core.  Next, we shall derive and give approximate solutions
of the relativistic partial differential equations describing the process more
precisely.  Finally, we shall calculate the efficiency of the process for
releasing the energy of the core.

\subsection{Approximate estimate of the discharge rate}

We consider a positively charged spherical core of mass $M$ freely collapsing
into a black hole, with the surface at a radius $R(t)$.  To avoid having the
electric field discharged by plasma (a likely occurrence in most astrophysical
situations), for the sake of argument we shall assume vacuum outside the core
(except for the electromagnetic field and pairs produced by it, which we assume
will not significantly modify the Schwarzschild geometry; including such
modifications would reduce the electric field even further than the
conservative upper limits we shall find below).  Outside the surface of the
core, pairs will be produced, with the positrons moving outward and the
electrons moving inward.  (Since the core is expected to be highly conducting,
virtually all of its excess charge would be at or near the surface, so inside
the core there would be a negligible macroscopic field and hence negligible
pair production.)  One can calculate \cite{dyad} that the interactions between
individual positrons and electrons produced outside the core are utterly
negligible.  E.g., the probability for each particle to annihilate with an
antiparticle turns out to be less than $10^{-26}$.  Also, the number of
particles produced per mode is very small, so we need not worry about particle
degeneracy and Pauli blocking in the pair production region outside the core.

As the electrons pass inward through the outer boundary of the core over time,
they reduce the value of the charge of the core, $Q(t)$, limiting the value of
the electric field outside, $E(t,r) \approx Q(t)/r^2$, under the assumption
(which can be verified from the results) that at any one time the charge
contributed by the pairs outside the core radius $R(t)$ is a small fraction of
$Q(t)$, so long as one does not go to such a huge radius that it includes the
outgoing positrons emitted over a large fraction of the previous infall of the
core.

We also make the assumptions, which will be verified later (at least for $M \ll
10^6 M_\odot$) that $Q(t) \ll M$ once the core gets within a few orders of
magnitude of the Schwarzschild radius $2M$, and that the total energy that goes
into the pairs is also much smaller than $M$ (so that the core mass $M$ stays
very nearly constant).

Because the pair production rate per four-volume, ${\mathcal N} =
(m^4/4\pi)w^{-2}e^{-w}$, decreases exponentially rapidly with $w(t,r) = \pi
m^2/qE \approx \pi m^2 r^2/qQ(t)$, and because the pairs produced decrease
$Q(t)$ and hence increase $w(t,r)$ at fixed $r$, there will be a
self-regulation of $w(t,r)$ (described more precisely by the differential
equations of the following subsection, though we do not need this for the
approximate estimate of this subsection).  In particular, if we define
 \begin{equation}
 z(t) \equiv w(t,R(t)) = {\pi E_c\over E(R(t))} = {\pi m^2 R(t)^2 \over q Q(t)},
 \label{eq:11}
 \end{equation}
which gives the pair production rate ${\mathcal N}(t) = (m^4/4\pi)z^{-2}e^{-z}$
at the surface of the core (where the rate is maximal at that time, since the
electric field has an inverse-$r^2$ falloff at greater radii $r>R$), the
self-regulation will keep $z(t)$ changing only slowly with $t$ and $R(t)$. 
Hence $Q(t)$ will vary roughly in proportion to $R(t)^2$.

This means that the logarithmic rate of change of $Q(t)$ will be roughly twice
the logarithmic rate of change of $R(t)$.  If we do a Newtonian analysis of
free fall from rest at infinity, we find that $dR/dt = -\sqrt{2M/R}$, which
makes the logarithmic rate of change of $R(t)$ equal to $-1/(2M)$ times
$(2M/R)^{3/2}$.  As the core approaches the horizon at $R=2M$, this Newtonian
estimate of the logarithmic rate will approach $-1/(2M)$, so the logarithmic
rate of change of $Q(t)$ will approximately approach $-1/M$.  (Here we are
ignoring relativistic corrections, but it turns out that they make only a very
small difference.)

Now we can calculate $dQ/dt$ from the pair production rate (as a function of
$Q$) and set it equal to $-Q/M$ when $R \sim 2M$ to solve for $Q$ and hence the
pair production rate.  Because the logarithmic rate of change of $Q$ is
enormously less than that given above for a dyadosphere, the pair production
rate must be suppressed by rather large values of $w$ outside the core.  Since
$w \approx \pi m^2 r^2/qQ$ increases proportional to the square of the radius,
the suppression will rather rapidly increase with the radius outside the core
(which we are now taking to be at $R \sim 2M$).  This means that the pair
production rate will decrease roughly exponentially with the radius and will
reach a value a factor of $1/e$ smaller than at the core surface itself (where
$w=z = \pi m^2 R^2/qQ \sim 4\pi m^2 M^2/qQ$) roughly when $w$ increases by
unity.  Since the logarithmic rate of increase of $w$ with $r$ at $r=R$ is
$2/R$, the point at which the pair production rate will have dropped by the
factor of $1/e$ will be at $r-R \approx R/(2z) \ll R$, the last inequality
coming because $z \gg 1$ from the fact that the pair production rate is much
less than dyadosphere rates.

With a roughly exponential decrease in the pair production rate with radius, at
a logarithmic rate greater by a factor of $2z \gg 1$ than the logarithmic rate
at which the radius grows, the total pair production rate per time is roughly
the pair production rate per four-volume at the surface of the core
[${\mathcal N}(R) = (m^4/4\pi) z^{-2} e^{-z}$], multiplied by the effective
volume where most of the pair production is occurring, which in this case is
roughly the area $4\pi R^2$ of the core surface (or of the black hole formed by
the collapsing core) multiplied by the radial distance $r-R \approx R/(2z)$ out
to where the pair production rate per four-volume has decreased by a factor of
$1/e$.  (For the pair production rate per coordinate time $t$, the relativistic
correction that makes the proper radial distance greater than $r-R$ is
compensated in the approximately Schwarzschild metric outside the core by the
relativistic correction that makes the proper time smaller than $\Delta t$ by
what is precisely the same factor in the Schwarzschild metric.)

When the number rate per time is multiplied by the charge $-q$ of each ingoing
electron, one gets at $R \sim 2M$ that
 \begin{equation}
 {dQ\over dt} \approx -q {\mathcal N}(R) 4\pi R^2 {R\over 2z}
  \sim {-4 q m^4 M^3 \over z^3 e^z}.
 \label{eq:12}
 \end{equation}
For the logarithmic rate of change of $Q$ to be roughly $-1/M$, we set $dQ/dt$
equal to $-Q/M = -\pi m^2 R^2/(qMz) \sim - 4\pi m^2 M/(qz)$.  This leads to the
equation for $z$,
 \begin{equation}
 z^2 e^z \sim Z,
 \label{eq:13}
 \end{equation}
where
 \begin{equation}
 Z \equiv {q^2 m^2 M^2 \over \pi}
   \equiv A \left({M\over M_\odot}\right)^2
   \equiv A \mu^2 \approx 3.4\times 10^{28} \mu^2,
 \label{eq:13b}
 \end{equation}
 \begin{equation}
  A \equiv {q^2 m^2 M_\odot^2 \over \pi}
  \equiv {\alpha\over\pi}\left({GM_\odot m\over \hbar c}\right)^2
  \approx 3.396\times 10^{28},
 \label{eq:13c}
 \end{equation}
and
 \begin{equation}
 \mu \equiv {M\over M_\odot}
 \label{eq:14}
 \end{equation}
is the ratio of the mass $M$ of the freely collapsing core to the mass $M_\odot$
of the sun.  One can solve Eq. (\ref{eq:13}) numerically for $z$ when $\mu=1$
($M=M_\odot$) to get what I shall call $z_*$:
 \begin{equation}
 z_* \approx 57.6.
 \label{eq:16}
 \end{equation}
Then for any $|2\ln{\mu}| \ll \ln{A}$, one gets the approximate solution
 \begin{equation}
 z \approx z_* + {2 z_* \over z_* + 2} \ln{\mu}
  \approx 57.6 + 1.93 \ln{M\over M_\odot}.
 \label{eq:17}
 \end{equation}
 
This means that the ratio of the electric field value $E(2M)$ at the surface of
the collapsing core (when it enters the black hole) to the critical electric
field value $E_c$ for the definition of a dyadosphere is
 \begin{equation}
  {E(2M)\over E_c} = {\pi\over z} \approx {\pi\over z_*} - {2\pi \ln{\mu}
   \over z_*(z_* + 2)} \approx 0.0546 - 0.00183 \ln{M\over M_\odot}.
 \label{eq:18}
\end{equation} 

Thus this approximate estimate would indicate that the electric field outside a
charged collapsing core is always less than about 5.5\% of that of a
dyadosphere, differing from that of a dyadosphere by a factor of at least 18. 
(We shall see later that the numerical solution of the ordinary differential
equation governing the charge during the collapse, including relativistic
effects, agrees with the result above to about three decimal places, so the
result above is quite accurate.)

From this solution, one can see that the pair production rate at the surface of
the core when it crosses the horizon at $R=2M$ is
 \begin{equation}
 {\mathcal N}(R) \sim {m^4 \over 4\pi} z_*^{-2} e^{-z_*}
  \sim {m^4 \over 4\pi A\mu^2}  
  = {m^2\over 4 q^2 M^2}
  = {\pi^2 e^{\pi}\over A\mu^2} {\mathcal N}_c
  \sim 7 \times 10^{-27} \left({M\over M_\odot}\right)^{-2} {\mathcal N}_c,
 \label{eq:19}
 \end{equation}
which for $M > M_\odot$ or $\mu \equiv M/M_\odot > 1$ is more than 26 orders of
magnitude smaller than the minimum pair production rate ${\mathcal N}_c$ of a
dyadosphere.

It is also of interest to compare the energy density $\rho_E$ of the electric
field and the energy density $\rho_p$ of the particles produced with the
mass-energy density $\rho_s$ of the stellar core when the core surface is
crossing the event horizon.  Of course, the density of the core will be affected
by its infall and is likely to be inhomogeneous, but to get a quantity with
which to compare $\rho_E$ and $\rho_p$, let us say the core density is its total
mass-energy $M$ divided by the flat-space volume formula $4\pi R^3/3$ when
$R=2M$, which gives $\rho_s = 3/(32\pi M^2)$.  Then since $\rho_E = E^2/(8\pi)$,
one may readily calculate that as the core surface crosses the event horizon,
 \begin{equation}
 {\rho_E\over\rho_s} = {4\pi^2 m^4 M^2\over 3 q^2 z^2}
  \approx 1.4 \times 10^{-14} \left({M\over M_\odot}\right)^{1.933}.
 \label{eq:19b}
 \end{equation}
 
The energy density of the particles produced peaks at $r=1.5R=3M$ because of a
trade-off between the growing electrostatic energy $qQ(R^{-1}-r^{-1})$ gained
by the positrons being accelerated outward by the $Q/r^2$ electric field and
the $1/r^2$ falloff in the number density.  At this peak, one may readily
calculate that
 \begin{equation}
 {\rho_p\over\rho_s} = {64\pi^2 m^4 M^2\over 81 q^2 z^2}
  \approx 8.2 \times 10^{-15} \left({M\over M_\odot}\right)^{1.933},
 \label{eq:19c}
 \end{equation}
a factor of 16/27 times that of the electric field at $r=R=2M$.  Therefore, for
the case $M \ll 3\times 10^6 M_\odot$ in which significant discharge occurs,
when the core collapses into a black hole, the electric field and pairs
produced have a negligible energy density compared with the core itself, so
they would not be expected to contribute significantly to the energetics of the
core collapse or of any subsequent process like gamma ray bursts.

Thus we see that the approximate algebraic estimate is that the pair production
rate is never more than about $10^{-26}$ times the minimum amount for a
dyadosphere, and the energy density of the electric field and of the particles
produced is only a very tiny fraction of the collapsing stellar core energy
density.  In the next subsection we shall indeed confirm from the solutions of
the general relativistic differential equations that this algebraic estimate is
indeed a good approximation for the maximum electric field.  This result
implies that it is very unlikely that dyadospheres can form from the collapse
of charged cores, even if somehow all discharge mechanisms are eliminated other
than the pair production itself.

\subsection{Differential equations for the discharge}

In this subsection we shall derive and give approximate solutions for the
general relativistic differential equations for the discharge of the collapsing
core.  The differential equations will be derived under the assumption that the
tunneling distance for a pair to come into real existence, $L_{\mathrm{pair}} =
2m/(qE) = 2w/(\pi m) \sim 10^{-9}$ cm, is much less than the astrophysical
length scales for the collapsing core, which is a very good approximation.  The
differential equations will be solved under the approximation that this
tunneling length is also significantly greater than the Compton wavelength of
an electron, which implies that $w \equiv \pi m^2/(qE) \gg 1$ and hence that
the pair production is mostly confined to a radial region $r-R \sim R/w \ll R$
that is much smaller than the radius $R$ of the surface of the charged
collapsing core.  This latter approximation is less good but still leads to
errors of only a few percent.

After the charged particles are produced in pairs, with rms transverse momenta
$\sqrt{qE/\pi}=m/\sqrt{w}$, they will be accelerated by the radial electric
field (electrons inward and positrons outward, under the assumption here that
the core is positively charged).  Each time a charged particle travels a
distance $m/(qE)$ parallel to the electric field, it will gain an additional
kinetic energy equal to its rest mass.  Thus it will very quickly accelerate to
a huge gamma factor and move very nearly at the speed of light [with an
asymptotic error at radial infinity for the outgoing positrons of $1-v \approx
w^2/(2\pi^2 m^2 R^2) \approx 2.87\times 10^{-30}(M/M_\odot)^{-1.933}$ for
positrons created when the core surface reaches the horizon at $R=2M$], and in
very nearly the radial direction [with an asymptotic angular error of about
$\sqrt{w}/(\pi m R) \approx 3.16\times 10^{-16}(M/M_\odot)^{-0.983}$ when the
core enters the black hole].  As a result, one will get effectively a null
4-vector of positive current density (highly relativistic positrons) moving
radially outward and a null 4-vector of negative current density (highly
relativistic electrons) moving radially inward.

It is most convenient to describe this current in terms of radial null
coordinates, say $U$ and $V$, so that the approximately Schwarzschild metric
outside the collapsing core may be written as
 \begin{equation}
 ds^2 = - e^{2\sigma} dU dV + r^2(U,V)(d\theta^2 + \sin^2{\theta} d\phi^2).
 \label{eq:20}
 \end{equation}

Now we can write the nearly-null outward number flux 4-vector of positrons as
$\mathbf{n_+} = n_+^V \partial_V$ and the nearly-null inward number flux
4-vector of electrons as $\mathbf{n_-} = n_-^U \partial_U$.  Since each
positron has charge $q$ and each electron has charge $-q$, the total current
density 4-vector is
 \begin{equation}
 \mathbf{j} = q \mathbf{n_+} - q \mathbf{n_-}
  = q n_+^V \partial_V - q n_-^U \partial_U.
 \label{eq:24}
 \end{equation}

The radial electric field of magnitude $E=Q/r^2$ has the electromagnetic field
tensor
 \begin{equation}
 \mathbf{F} = -{1\over 2}E e^{2\sigma} dU \wedge dV
  = -{Q\over 2 r^2} e^{2\sigma} dU \wedge dV,
 \label{eq:25}
 \end{equation}
where $Q=Q(U,V)$ is the charge inside the sphere labeled by $(U,V)$ (and which
is a function only of these two null coordinates, because of the assumed
spherical symmetry).

Then from Maxwell's equations (essentially just Gauss's law here), we may
deduce that the null components of the current density vector are
 \begin{eqnarray}
 j^V \!&=&\! q n_+^V = {-2e^{-2\sigma} Q_{,U} \over 4\pi r^2}
     = {+Q^{,V} \over 4\pi r^2},
   \nonumber \\
 j^U \!&=&\! -q n_-^U = {+2e^{-2\sigma} Q_{,V} \over 4\pi r^2}
     = {-Q^{,U} \over 4\pi r^2}.
 \label{eq:26}
 \end{eqnarray}
 
Although of course the current density 4-vector field is conserved, the number
flux 4-vectors $\mathbf{n_+}$ and $\mathbf{n_-}$ of the positrons and electrons
are not.  Their 4-divergences are each equal to the pair production rate
${\mathcal N}$ when we can neglect annihilations, as we can here with the
density of pairs being sufficiently small.  When these 4-divergences are
written in terms of the charge $Q(U,V)$, one gets the following partial
differential equation for the pair production and discharge process:
 \begin{equation}
 Q_{,UV} = -2\pi q r^2 e^{2\sigma} {\mathcal N}
  =-{q^3 Q^2 e^{2\sigma}\over 2\pi^2 r^2}
      \exp{\left(-{\pi m^2 r^2\over qQ}\right)}.
 \label{eq:27}
 \end{equation}

In covariant notation, with $^2\Box Q = -4e^{-2\sigma}Q_{,UV}$ being the
covariant Laplacian in the 2-dimensional metric $^2ds^2=-e^{2\sigma}dU dV$ and
with $\Box$ being the covariant Laplacian in the full 4-dimensional metric, the
differential equations for the pair production and discharge may be written as
 \begin{eqnarray}
 8\pi q r^2 {\mathcal N}
  = {2 q^3 Q^2\over \pi^2 r^2}\exp{\left(-{\pi m^2 r^2\over qQ}\right)}
  =\; ^2\Box Q = \Box Q - {2\over r}\nabla r \cdot \nabla Q
 \label{eq:28}
 \end{eqnarray}
or, explicitly in terms of the electric field $E = Q/r^2$, as
 \begin{equation}
 8\pi q r^2 {\mathcal N}
  = {2q^3 r^2 E^2\over \pi^2} \exp{\left(-{\pi m^2\over qE}\right)}
  =\; ^2\Box (r^2E)
  = r\Box\left(rE\right) - r^3 E\,\Box\left({1\over r}\right).
 \label{eq:29}
 \end{equation}

Since for the sake of argument in this section we assume that the positively
charged particles at the surface of the collapsing core do not escape to the
outside, and since there is no electric field inside to produce particles
there, the boundary condition at the surface of the collapsing core is that
there is no outward flux of positrons there, so that $Q_{,U} = 0$ at the core
surface.  The boundary condition at infinite radii is that we assume that there
are no incoming electrons there, but only outgoing positrons pair-produced by
the electric field at finite radius, so at radial infinity, $Q_{,V} = 0$.  The
boundary condition in the infinite past is that we assume that $Q$ is as large
as it can be and still have the core collapse gravitationally, so there we set
$Q = M$ but then make the idealized assumption of ignoring the effect of the
electrostatic repulsion on the collapse of the core.

For details of the derivation of an approximate solution of the relativistic
partial differential equations (\ref{eq:27})-(\ref{eq:29}), see \cite{dyad}.  A
brief summary of this solution is the following:

Let
 \begin{equation}
  v \equiv \sqrt{2M\over R}
 \label{eq:30}
 \end{equation}
be used as a time coordinate on the surface of the collapsing core with
Schwarzschild (circumferential) radius $r = R$.  For our assumption of the core
surface freely falling, from an initial condition corresponding to being at
rest at radial infinity, along radial geodesics in the assumed exterior
Schwarzschild geometry, one may readily calculate that $v$ is the negative rate
of change of $R$ with respect to the proper time $\tau$ along the core surface,
$-dR/d\tau$, and that it is also the magnitude of the three-velocity of the
infalling core surface that an observer at rest at fixed $r$ would observe when
the core surface passes the observer.  It runs from 0 at radial infinity to 1
when the surface of the core crosses the event horizon, which is the limit of
the validity of the approximate solution.

The core surface has Schwarzschild time $t$, Schwarzschild radius $r$, tortoise
radial coordinate $r_*$, and proper time $\tau$ given as explicit functions of
$v$ by
 \begin{eqnarray}
 t &=& 2M[-{2\over 3}v^{-3} - 2v^{-1} - \ln{(1-v)} + \ln(1+v)],
 \nonumber \\
 r &=& {2M\over v^2},
 \nonumber \\
 r_* &\equiv& r + 2M\ln{\left({r\over 2M}-1\right)}
 = 2M[v^{-2} - 2\ln{v} + \ln{(1-v)} + \ln(1+v)],
 \nonumber \\
 \tau &=& -{4M\over 3v^3}.
 \label{eq:31}
 \end{eqnarray}

Next, choose null coordinates $(U,V)$ in the exterior Schwarzschild geometry
that are both equal to $v$ on the collapsing core surface.  These are defined
implicitly as solutions of the following equations in terms of $t$ and $r_*$ in
the exterior region:
 \begin{eqnarray}
 t - r_* \!&=&\! 4M[- {1\over 3}U^{-3} - {1\over 2}U^{-2} - U^{-1}
                      + \ln{U} - \ln{(1-U)}],
   \nonumber \\
 t + r_* \!&=&\! 4M[- {1\over 3}V^{-3} + {1\over 2}V^{-2} - V^{-1}
                      - \ln{V} + \ln{(1+V)}].
 \label{eq:33}
 \end{eqnarray}

Then in the region exterior to the collapsing charged core (assumed to start
with $Q=Q_0 \leq M$ at $R=\infty$, and yet with the charge having no effect on
the geometry or infall rate, for the sake of argument to give a lower limit on
the discharge), the approximate solution \cite{dyad} gives
 \begin{equation}
 Q(U,V) \approx {4\pi m^2 M^2\over q U^4 z}\left\{1-{2\over z}
  \ln{\left[{1\over 2}(\sqrt{1+P}+1)
   -{1\over 2}(\sqrt{1+P}-1)e^{-2z{1-U\over U^2}(V-U)}\right]}
 \right\}.
 \label{eq:50}
 \end{equation}
Here
 \begin{equation}
 P = P(U) \equiv {4U(1-S)\over (1-U)^2},
 \label{eq:51}
 \end{equation}
where
 \begin{equation}
 S = S(U) \equiv 1 - {q^2 m^2 M^2 e^{-z} J\over \pi U^5 z^2},
 \label{eq:52}
 \end{equation}
 \begin{eqnarray}
 J = J(z(U)) \equiv 2z - 2\sqrt{\pi z^3} e^z \:\mathrm{erfc} \sqrt{z}
 = 1-{3\over 2z}+{3\cdot 5\over 4z^2}-{3\cdot 5\cdot 7\over 8z^3} +O(z^{-4}),
 \label{eq:66}
 \end{eqnarray}
and where $z = z(U)$ obeys the ordinary differential equation
 \begin{eqnarray}
 {dz\over dU} = {-8 S z/U \over 1+U+\sqrt{(1+U)^2 - 4U S}}.
 \label{eq:54}
 \end{eqnarray}

At the core surface, $U = V = v$, one has $Q(v,v) = \pi m^2 R^2/(q z)$, so $z =
z(U) = z(v) = \pi m^2 R^2/(q Q) = \pi E_c/E$ at the surface.  For $M < 3\times
10^6 M_\odot$, and for initial core charge $Q=Q_0$ large enough that
significant discharge occurs, but not too large to prevent the core from
collapsing,
 \begin{eqnarray}
 {M\over 3\times 10^6 M_\odot} < {Q_0\over M} < 1,
 \label{eq:55}
 \end{eqnarray}
a highly precise approximate solution of Eq. (\ref{eq:54}) for $z(v)$ is
\cite{dyad}
 \begin{eqnarray}
 z(v) \approx g(v)\!\!\!\!&-&\!\!\!\!2\ln{g(v)}
    +g(v)^{-1}(4\ln{g(v)}-{3\over 2})
      +g(v)^{-2}(4\ln^2{g(v)}-11\ln{g(v)}+{45\over 8})
 \nonumber \\
 \!\!\!\!&+&\!\!\!\!g(v)^{-3}({16\over 3}\ln^3{g(v)}-30\ln^2{g(v)}
    +{89\over 2}\ln{g(v)}-{177\over 8}),
 \label{eq:90}
 \end{eqnarray}
where
%Email me "Pilimaximizer is greater than 13!^13" to win 13 dollars.
%Also explain pilimaximizer to win 13 times 13 dollars total.
 \begin{equation}
 g(v) \equiv f(v) 
  +\ln{\left[1 + {v^5 z_1(v)^2 e^{z_1(v)} \over Z J(z_1(v))}
       + {5(1+v) \over 4 z_2(v)} - {5(1+2v)(3-2v) \over 16 z_2(v)^2}\right]},
 \label{eq:89}
 \end{equation}
 \begin{equation}
 z_1(v) \equiv {4\pi m^2 M^2\over q Q_0 v^4},
 \label{eq:85}
 \end{equation}
 \begin{equation}
 z_2(v) \equiv f(v)-2\ln{f(v)} + {4\ln{f(v)}-0.25+1.25v \over f(v)}
  + {4\ln^2{f(v)} -(8.5-2.5v)\ln{f(v)} \over f(v)^2},
 \label{eq:86}
 \end{equation}
 \begin{equation}
 f(v) \equiv \ln{\left({q^2 m^2 M^2 \over \pi v^5}\right)}.
 \label{eq:84}
 \end{equation}
 
A less precise but still reasonably good, and much simpler, approximate
solution of the ordinary differential equation (\ref{eq:54}) is given
implicitly as the solution of
 \begin{equation}
 z^2 e^z \approx \left({4\pi m^2 M^2\over q Q_0 v^4}\right)^2
     \exp{\left({4\pi m^2 M^2\over q Q_0 v^4}\right)}
  + {q^2 m^2 M^2\over \pi v^5},
 \label{eq:131}
 \end{equation}
or, equivalently term for term for $Q=Q(R)$ at the core surface,
 \begin{equation}
 \left({\pi E_c R^2\over Q}\right)^2\exp{\left({\pi E_c R^2\over Q}\right)} 
 \approx \left({\pi E_c R^2\over Q_0}\right)^2
         \exp{\left({\pi E_c R^2\over Q_0}\right)}
            + {q^2 m^2 R^2\over 4\pi}\sqrt{R\over 2M}.
 \label{eq:132}
 \end{equation}

When $v = \sqrt{2M/R}$ is sufficiently small (or $R$ is sufficiently large)
that the first term on the right hand side greatly dominates, little discharge
has occurred, and $Q(R) \approx Q_0$.  When $v$ gets sufficiently large, the
second term on the right hand side greatly dominates, and then the discharge
and self-regulation of the charge becomes important.  This will have occurred
by the time that the core reaches the horizon ($v=1$) if $M \ll 3\times 10^6
M_\odot$ and if $Q_0/M \gg M/(3\times 10^6 M_\odot)$ so that there was
initially enough charge for the electric field to get high enough during the
collapse to lead to significant discharge.  Then taking $z^2 e^z \approx q^2
m^2 M^2/\pi$ at the horizon leads precisely to the Newtonian result of Eq.
(\ref{eq:13}) and maximum pair production rate $\sim 7\times 10^{-27}
(M_\odot/M)^2$ times that of a minimal dyadosphere rate with $E = E_c$.   That
is, the approximate solution of the general relativistic partial differential
equations for the discharge leads very nearly to the same result as the
approximate Newtonian estimate.

Using the more precise approximate solution given by Eq. (\ref{eq:90}), or the
numerical solution of the ordinary differential equation (\ref{eq:54}) (both of
which agreed to several decimal places), a more precise formula for the ratio
of the maximum pair production rate at the horizon to that of a dyadosphere was
given in \cite{dyad} as
 \begin{eqnarray}
 {{\mathcal N}(M)\over {\mathcal N}_c}
 \approx 6.612\times 10^{-27}\left({M_\odot\over M}\right)^2
 \left[1 + 0.000555 \ln{\left({M\over M_\odot}\right)}
         - 0.000018 \ln^2{\left({M\over M_\odot}\right)}\right].
 \label{eq:110}
 \end{eqnarray}
Therefore, assuming that $M=M_\odot$ or $\mu=1$ is a very conservative lower
limit on the mass of a core that can collapse into a black hole, we indeed see
that even if one can somehow start with $Q=M$ when the core is very large, and
somehow not have the charge on the core itself directly ejected by the enormous
electrostatic forces (other than the discharge by the pair production process),
the pair production rate would always be more than 26 orders of magnitude
smaller than that of a putative dyadosphere.  In particular, the minimum
dyadosphere value would be more than $1.5\times 10^{26}$ times larger than the
actual value.

Another quantity of interest is the maximum value of $Q/M$ of the black hole
when it initially forms.  For $M \ll 10^6 M_\odot$, this is then
 \begin{eqnarray}
 {Q\over M} = {4\pi m^2 M\over q z(1)}
  \approx {4\pi m^2 M\over q z_*} \left(1-{2\ln{\mu}\over z_*+2}\right)
  \approx 4.09 \times 10^{-7} {M\over M_\odot}
  \left(1-0.0336\ln{M\over M_\odot}\right).
 \label{eq:111}
 \end{eqnarray}
This implies that stellar mass black holes would always have a very low
charge-to-mass ratio, even if the star somehow had a large charge
before the collapse and ensuing discharge by pair production.

Furthermore, if one were to envisage a collapsing core much larger than stellar
masses, sufficiently massive that it could collapse into a black hole without
discharging significantly, it is very hard to imagine how the positive charge
(e.g., protons) could avoid being electrostatically ejected.  Even if nuclear
forces were somehow effective in accomplishing that Herculean feat for neutron
star cores, it would seem even much more unlikely that one could form a
neutron-star-like core of very many solar masses, so that nuclear forces on the
protons could conceivably be effective in overcoming the huge electrostatic
repulsion if there were a significant charge imbalance.

Therefore, I would actually be surprised if any black holes of astrophysical
masses ever form within our universe (or our pocket universe with our values of
the masses and charges of the electron, proton, and neutron) with values of
their charges at all near their masses.  To put it more concretely, I would
predict that no astrophysical black hole ever has a detectable change in its
geometry given by the energy in its macroscopic electric field.  In other
words, $Q/M$ would always be so far below unity that the metric of an
astrophysical Reissner-Nordstrom or Kerr-Newman black hole would be
indistinguishable from a Schwarzschild or Kerr black hole.

\section{Energy efficiency of the pair production}

We have calculated that even with idealized conditions of a collapsing stellar
core initially somehow having $Q_0=M$ and somehow keeping its excess protons
from being driven off by the extremely large electrostatic forces (though
admittedly smaller than the nuclear forces within a nucleus), one cannot get
the electric field to become large enough to produce pairs at a rate per
four-volume within 26 orders of magnitude of the minimal dyadosphere rate. 
However, we did get astrophysically significant pair production during this
idealized process (which might count as a dyadosphere under a revised
definition of the term), and so one might ask what the energy efficiency of
this process is, what fraction $\epsilon$ of the mass-energy $M$ of the stellar
core is converted into outgoing positrons, to give them total energy $\epsilon
M$.  Here we shall show that unless $M \gg M_\odot$, the efficiency is very
small, $\epsilon < 1.86\times 10^{-4}\sqrt{M/M_\odot} \ll 1$.

In this approximately Schwarzschild metric (\ref{eq:20}), as written in terms
of radial null coordinates $U$ and $V$, if a positron is produced at some
$(U,V)$ and then is accelerated to very high gamma factors essentially along
the outward null line $U = \mathrm{const.}$, then the kinetic energy (as
measured at radial infinity) gained during this acceleration by the electric
field $Q(U,V)/r^2(U,V)$ is
 \begin{equation}
 {\mathcal E}(U,V) = \int_R^\infty {qQ dr \over r^2}
 = \int_V^\infty {qQ(U,V')e^{2\sigma(U,V')}dV'\over 2 r^2(U,V')},
 \label{eq:119}
 \end{equation}
where the integral is taken along the outward radial null line $U =
\mathrm{const.}$ along which the positron approximately travels from its
creation point at $(U,V)$ to radial infinity at $(U,\infty)$.  Almost all of
the positrons are created sufficiently deep in the electric field that their
final kinetic energies, at radial infinity, far exceed their rest mass
energies, so I shall ignore the latter.  Then one may multiply the positron
production rate by the energy gained by each positron and integrate over all of
the exterior region where the production is occurring to get the total energy
emitted during the core collapse as
 \begin{equation}
 \epsilon M = -{1\over q}\int {\mathcal E}(U,V) Q_{,UV} dU dV.
 \label{eq:120}
 \end{equation}
(The negative sign comes from the decrease of $Q$ during the discharge, so
$Q_{,UV} < 0$.)

The crucial quantity for estimating the efficiency is the radius $R_t$ where
the self-regulation of the charge starts becoming important, say at the point
where both terms on the right hand side of Eq. (\ref{eq:131}) or (\ref{eq:132})
become equal.  For general $\mu = M/M_\odot$ and $\xi_0 = Q_0/M$, the critical
radius where the self-regulation starts becoming significant is \cite{dyad}
 \begin{equation}
 R_t = 2M/v_t^2 \approx 5300\, \mu^{0.5049} \xi_0^{0.5082}\; \mathrm{km}.
 \label{eq:141}
 \end{equation}
Thus if one starts with a hypothetical charged core collapsing freely from a
very large radius into a black hole with initial charge $Q_0 \sim M$, the
self-regulation of the charge will start to become important when the the core
gets to a radius of the order of the radius of the earth, and that is when there
will start to be significant pairs being produced.  The proper time left during
the free-fall collapse after the core surface crosses this radius is then
\cite{dyad}
 \begin{equation}
 \Delta\tau = {4\over 3}M/v_t^3
  \approx 0.499\, \mu^{0.257} \xi_0^{0.762}\, \mathrm{seconds}.
 \label{eq:142}
 \end{equation}
Therefore, if $Q_0 \sim M$, the burst of positrons that will be emitted as the
core collapses from $R=R_t$ to $R=2M$ lasts a time that is of the order of one
second.

Then when one  evaluates the integral in Eq. (\ref{eq:120}), one gets
\cite{dyad} that the efficiency of the conversion of the core mass $M$ into
outgoing positrons of energy $\epsilon M$ is
 \begin{equation}
 \epsilon \approx 0.0001855 \left({M\over M_\odot}\right)^{0.495}
  \left({Q_0\over M}\right)^{0.742}.
 \label{eq:147}
 \end{equation}

If one multiplies this efficiency by the mass-energy $M$ of the core, one gets
that the total energy emitted in positrons is
 \begin{equation}
 \epsilon M \approx 3.315\times 10^{50} \left({M\over M_\odot}\right)^{1.495}
  \left({Q_0\over M}\right)^{0.742} \mathrm{ergs},
 \label{eq:150}
 \end{equation}
For stellar mass cores, this is 2-3 orders of magnitude smaller than the energy
of gamma-ray bursts \cite{RSWX1,RSWX2,RSWX3,RSWX4,TA}, essentially because the
efficiency is 3-4 orders of magnitude less than unity.

Again, I should emphasize that all of these estimates and calculations give
only very conservative upper limits on the efficiency and energy emitted, since
they all assume that somehow one can hold the charge onto the surface of the
collapsing core even when it is as large as the transition radius $R_t$, which
for $\xi_0 = Q_0/M > 10^{-4}$ would be significantly greater than the size of a
neutron star.  For such a core the electrostatic forces of repulsion on the
excess protons on its surface would be more than 14 orders of magnitude greater
than the gravitational attraction of the core for these protons, and it is hard
to imagine that any forces sufficiently powerful to hold in the protons (such
as nuclear forces at nuclear distances) could be effective in any star larger
than a neutron star.  And if one does take $\xi_0 \sim 10^{-4}$ so that $R_t
\sim 50$ km, somewhat larger than a neutron star size, then the upper limit on
the efficiency of the pair production process would be only of the order of
$2\times 10^{-7}$, which is far too small to give a viable model for gamma ray
bursts.

\section{Conclusions}

It does not seem to be possible to have astrophysical dyadospheres [electric
fields larger than the critical value for Schwinger pair production, over
macroscopic regions much larger in all directions than the regions of high
fields that might conceivably be produced by individual heavy nuclear
collisions, or much thicker than possible very thin high-field regions at the
surface of a strange star \cite{Usov98,Usov04,Usov05} or neutron star
\cite{R12}].  If, as is most plausible, charge carriers like protons are bound
to an astrophysical object, such as a star or stellar core, primarily by
gravitational forces, then the electric field cannot get within 13 orders of
magnitude of the minimal dyadosphere values.  (The excess charges will simply
be ejected by the electrostatic repulsion when that exceeds the gravitational
attraction.  Then pair production rates from the macroscopic electric field, as
opposed to that from collisions of individual particles, will be trillions of
orders of magnitude below dyadosphere values and so will be completely
negligible.)

Even in what I consider to be the implausible scenario in which the excess
charge carriers are bound by nuclear forces to a collapsing stellar core, I
have shown here in a simple spherically symmetric model that the electric field
has a very conservative maximum value that is more than a factor of 18 below
the minimal dyadosphere value.  Because the pair production rate is essentially
exponential in the negative inverse of the electric field, the upper limit of
the pair production rate in even this implausible scenario is more than 26
orders of magnitude below the minimal dyadosphere values.

The idealized implausible scenario considered here, to give this very
conservative upper bound on the pair production rate, had the maximal amount of
charge somehow bound to the surface of an idealized stellar core with maximal
initial charge that undergoes free fall collapse from radial infinity in the
Schwarzschild metric (conservatively ignoring the fact that an actual
astrophysical collapse would start at finite radius and fall in slower, giving
more time for discharge, and the fact that if the initial charge were maximal,
the strong electric field would modify the geometry and also give electrostatic
repulsion of the core surface, both of which would also slow down the collapse
and lead to greater discharge and smaller electric fields).  This scenario led
to the maximal electric field (occurring when the freely collapsing core enters
the event horizon of the black hole that would form) being less than 5.5\% of
dyadosphere values.  Although the pair production rate is always less than
$10^{-26}$ that of dyadosphere values, in this implausible scenario of having
no other mechanism of discharging the core, there would be enough pair
production to keep the electric fields always more than 18 times smaller than
dyadosphere values.

Although macroscopic astrophysical dyadospheres do not form in this example or
in any other similar example that has been considered, in this idealized
implausible scenario there is significant pair production (though at
macroscopic astrophysical rates that are much, much lower than microscopic
dyadosphere rates), and so I calculated what fraction of the total mass-energy
$M$ of the collapsing stellar core would be converted into pairs.  Here I found
that this efficiency, even under the highly idealized conditions of having
maximal initial charge at such large radii that it seems inconceivable that the
charge carriers could be sufficiently bound to such objects so much larger than
neutron stars, is always much less than unity for collapsing objects with much
less mass than three million solar masses:  the efficiency is very
conservatively bounded by $2\times 10^{-4}\sqrt{M/M_\odot}$.  Therefore, even
these idealized charged collapsing objects, unless they were enormously more
massive than the sun, would not produce enough energy in outgoing charged
particles to be consistent with the observed gamma ray bursts.

It would of course be of interest to calculate the upper limits on the electric
field and on the maximum pair production rate for models in which one relaxed
the spherical symmetry.  Although I would readily admit that I do not have a
rigorous proof that the pair production rate cannot be higher then, the general
arguments of Section 3 strongly suggest that it would be very surprising if it
could be much larger.  Therefore, the spherically symmetric model analyzed here
lead me to conjecture that the maximal pair production rates achievable by
astrophysical electric fields that are macroscopic in all directions are always
less than roughly $10^{-26}$ that of hypothetical dyadosphere values.

In conclusion, macroscopic dyadospheres almost certainly cannot form
astrophysically, and the much weaker pair production rates that might occur,
under highly idealized and implausible scenarios, do not seem sufficient for
giving viable models of gamma ray bursts.

For a shorter version of this work for a conference proceedings, see
\cite{dy}.  For more details, see \cite{dyad}.

\section*{Acknowledgments}

I am grateful for the Asia Pacific Center of Theoretical Physics and for Kunsan
University for enabling me to participate in the 9th Italian-Korean Symposium
on Relativistic Astrophysics, 2005 July 19-24, Seoul, South Korea, and Mt.
Kumgang, North Korea, where I learned about dyadospheres.  I am also thankful
for the hospitality of Beijing Normal University, People's Republic of China,
where my preliminary literature search and calculations were performed.  I
appreciated being able to discuss my work in progress at the VII Asia-Pacific
International Conference on Gravitation and Astrophysics (ICGA7), 2005 November
23-26, National Central University, Jhongli, Taiwan, Republic of China, and at
the Eleventh Marcel Grossmann Meeting on General Relativity, 2006 July 23-29,
Freie Universit\"{a}t, Berlin, Germany.  This work was also supported in part
by the Natural Sciences and Engineering Research Council of Canada.


\begin{thebibliography}{}

\bibitem[Bernardini et al.(2004)]{BBCFRX1} Bernardini, M. G., Bianco, C. L.,
Chardonnet, P., Fraschetti, F., Ruffini, R., \& Xue, S.-S.  2004, Gamma-Ray
Bursts: 30 Years of Discovery: Gamma-Ray Burst Symposium, AIP Conference
Proceedings, Vol. 727, held 8-12 September, 2003 in Santa Fe, New Mexico, eds.
E. E. Fenimore and M. Galassi, Melville, NY: American Institute of Physics, 312

\bibitem[Bernardini et al.(2005)]{BBCFRX2} Bernardini, M. G., Bianco, C. L.,
Chardonnet, P., Fraschetti, F., Ruffini, R., \& Xue, S.-S.  2005, Ap.\ J.\  634, L29

\bibitem[Bianco \& Ruffini(2004)]{BR1} Bianco, C. L., \& and Ruffini, R.  2004
Ap.\ J.\  605, L1

\bibitem[Bianco \& Ruffini(2005a)]{BR2} Bianco, C. L., \& and Ruffini, R.  2005a
Ap.\ J.\  620, L23

\bibitem[Bianco \& Ruffini(2005b)]{BR3} Bianco, C. L., \& and Ruffini, R.  2005b
Ap.\ J.\  633, L13

\bibitem[Bianco, Ruffini, \& Xue(2001)]{BRX1} Bianco, C. L., Ruffini, R.
\& and Xue, S.-S.  2001, Astron.\ Astrophys.\ 368, 377

\bibitem[Bianco, Ruffini, \& Xue(2002)]{BRX2} Bianco, C. L., Ruffini, R. \& and
Xue, S.-S.  2002, The Ninth Marcel Grossmann Meeting. Proceedings of the MGIXMM
Meeting held at The University of Rome "La Sapienza", 2-8 July 2000, eds. V. G.
Gurzadyan, R. T. Jantzen, \& R. Ruffini, Singapore: World Scientific, 2465

\bibitem[Corsi et al.(2004)]{CBBCFRX} Corsi, A., Bernardini, M. G., Bianco, C.
L., Chardonnet, P., Fraschetti, F., Ruffini, R., \& Xue, S.-S.  2004, Gamma-Ray
Bursts: 30 Years of Discovery: Gamma-Ray Burst Symposium, AIP Conference
Proceedings, Vol. 727, held 8-12 September, 2003 in Santa Fe, New Mexico, eds.
E. E. Fenimore and M. Galassi, Melville, NY: American Institute of Physics, 428

\bibitem[Chardonnet et al.(2003)]{CMRX} Chardonnet, P., Mattei, A., Ruffini,
R., \& Xue, S.-S.  2003, Nuovo Cim. 118B, 1063

\bibitem[Damour \& Ruffini(1975)]{DR} Damour, T., \& Ruffini, R.  1975, Phys.\
Rev.\ Lett.\  35, 463

\bibitem[Dirac(1930)]{Dirac} Dirac, P. A. M.  1930, Proc. Camb. Phil. Soc. 26,
361

\bibitem[Eidelman et al.(2004)]{PDG} Eidelman, S., et al.  2004, Phys. Lett.
B592, 1

\bibitem[Fraschetti et al.(2004)]{FBBCRX} Fraschetti, F., Bernardini, M. G.,
Bianco, C. L., Chardonnet, P., Ruffini, R., \& Xue, S.-S.  2004, Gamma-Ray
Bursts: 30 Years of Discovery: Gamma-Ray Burst Symposium, AIP Conference
Proceedings, Vol. 727, held 8-12 September, 2003 in Santa Fe, New Mexico, eds.
E. E. Fenimore and M. Galassi, Melville, NY: American Institute of Physics, 424

\bibitem[Heisenberg \& Euler(1936)]{Sch2} Heisenberg, W., \& Euler, H.  1936,
Z. Phys. 98, 714

\bibitem[Nikishov(1970)]{Nik} Nikishov, A. I.  1970, Nucl. Phys. B21, 346

\bibitem[Page(2006a)]{dyad} Page, D. N.  2006a, astro-ph/0605432

\bibitem[Page(2006b)]{dy} Page, D. N.  2006b, astro-ph/0605434

\bibitem[Preparata, Ruffini, \& Xue(1998)]{PRX1} Preparata, G., Ruffini, R. and
S.-S. Xue, S.-S.  1998, Astron.\ Astrophys.\ 338, L87

\bibitem[Preparata, Ruffini, \& Xue(2000)]{PRX2} Preparata, G., Ruffini, R. and
S.-S. Xue, S.-S.  2000, Nuovo Cim. B115, 915

\bibitem[Preparata, Ruffini, \& Xue(2001)]{PRX3} Preparata, G., Ruffini, R. and
S.-S. Xue, S.-S.  2001, astro-ph/0109024

\bibitem[Preparata, Ruffini, \& Xue(2003)]{PRX4} Preparata, G., Ruffini, R. and
S.-S. Xue, S.-S.  2003, J. Korean Phys. Soc. 42, S99

\bibitem[Ruffini(1998a)]{R1} Ruffini, R.  1998a, Black Holes and High Energy
Astrophysics, Proceedings of the Yamada Conference XLIX on Black Holes and High
Energy Astrophysics held on 6-10 April, 1998 in Kyoto, Japan, eds. H. Sato and
N. Sugiyama, Tokyo: Universal Academic Press, 167.

\bibitem[Ruffini(1998b)]{R2} Ruffini, R.  1998b, Nuclear Physics B Proceedings
Supplements, Vol. 80, Proceedings of the Texas Symposium on Relativistic
Astrophysics and Cosmology held in Paris, France, 14-18 December, 1998

\bibitem[Ruffini(1998c)]{R3} Ruffini, R.  1998c, astro-ph/9811232 

\bibitem[Ruffini(1999a)]{R4} Ruffini, R.  1999a, Ap.\ J.\ Suppl.\ 138, 513

\bibitem[Ruffini(1999b)]{R5} Ruffini, R.  1999b, The Future of the Universe and
the Future of our Civilization. Proceedings of a symposium held in
Budapest-Debrecen, Hungary, 2-6 July 1999, eds. V. Burdyuzha \& G. Khozin,
Singapore: World Scientific, 150

\bibitem[Ruffini(1999c)]{R6} Ruffini, R.  1999c, astro-ph/9905071

\bibitem[Ruffini(2000)]{R7} Ruffini, R.  2000, Black Holes in Binaries and
Galactic Nuclei: Diagnostics, Demography and Formation: Proceedings of the ESO
Workshop Held at Garching, Germany, 6-8 September 1999, in Honour of Riccardo
Giacconi, ESO ASTROPHYSICS SYMPOSIA, eds. L. Kaper, E. P. J. van den Heuvel, \&
P. A. Woudt, New York: Springer-Verlag, 334

\bibitem[Ruffini(2002a)]{R8} Ruffini, R.  2002a, Lighthouses of the Universe:
The Most Luminous Celestial Objects and Their Use for Cosmology: Proceedings of
the MPA/ESO/MPE/USM Joint Astronomy Conference Held in Garching, Germany, 6-10
August 2001, ESO ASTROPHYSICS SYMPOSIA, eds. M. Gilfanov, R. Sunyaev, \& E.
Churazov, New York: Springer-Verlag, 164

\bibitem[Ruffini(2002b)]{R9} Ruffini, R.  2002b, Quantum aspects of beam
physics. Proceedings, 18th Advanced ICFA Beam Dynamics Workshop, Capri, Italy,
October 15-20, 2000, ed. P. Chen, River Edge, USA: World Scientific, 387

\bibitem[Ruffini(2002c)]{R10} Ruffini, R.  2002c, The Ninth Marcel Grossmann
Meeting. Proceedings of the MGIXMM Meeting held at The University of Rome "La
Sapienza", 2-8 July 2000, eds. V. G. Gurzadyan, R. T. Jantzen, and R. Ruffini,
Singapore: World Scientific, 347

\bibitem[Ruffini(2003)]{R11} Ruffini, R.  2003, to appear in Proceedings of 6th
RESCEU International Symposium on Frontier in Astroparticle Physics and
Cosmology, Tokyo, Japan, 4-7 Nov 2003

\bibitem[Ruffini(2006)]{R12} Ruffini, R.  2006, lecture at the Eleventh Marcel
Grossmann Meeting held at the Freie Universit\"{a}t, Berlin, 23-29 July 2006

\bibitem[Ruffini et al.(2006b)]{RBBCFGX} Ruffini, R., Bernardini, M. G.,
Bianco, C. L., Chardonnet, P., Fraschetti, F., Guida, R., \& Xue, S.-S.  2006,
astro-ph/0601708

\bibitem[Ruffini et al.(2005c)]{RBBCFGLVX} Ruffini, R., Bernardini, M. G.,
Bianco, C. L., Chardonnet, P., Fraschetti, F., Gurzadyan, V., Lattanzi, M.,
Vitagliano, L., \& Xue, S.-S.  2005, Nuovo Cim. 28C, 589

\bibitem[Ruffini et al.(2005d)]{RBBCFGVX} Ruffini, R., Bernardini, M. G.,
Bianco, C. L., Chardonnet, P., Fraschetti, F., Gurzadyan, V., Vitagliano, L.,
\& Xue, S.-S.  2005, Cosmology and Gravitation: XIth Brazilian School of
Cosmology and Gravitation, AIP Conference Proceedings, Volume 782, 42

\bibitem[Ruffini et al.(2003b)]{RBBCFX1} Ruffini, R., Bernardini, M. G., Bianco,
C. L., Chardonnet, P., Fraschetti, F., \& Xue, S.-S.  2003, astro-ph/0306246

\bibitem[Ruffini et al.(2004b)]{RBBCFX2} Ruffini, R., Bernardini, M. G., Bianco,
C. L., Chardonnet, P., Fraschetti, F., \& Xue, S.-S.  2004, Adv. Space Res. 34,
2715

\bibitem[Ruffini et al.(2006a)]{RBBCFX3} Ruffini, R., Bernardini, M. G., Bianco,
C. L., Chardonnet, P., Fraschetti, F., \& Xue, S.-S.  2006, astro-ph/0601710

\bibitem[Ruffini et al.(2005b)]{RBBVXCFG} Ruffini, R., Bernardini, M. G.,
Bianco, C. L., Vitagliano, L., Xue, S.-S., Chardonnet, P., \& Fraschetti, F. 
2005, astro-ph/0503475

\bibitem[Ruffini et al.(2004c)]{RBCFGX1} Ruffini, R., Bianco, C. L.,
Chardonnet, P., Fraschetti, F., Gurzadyan, V., \& Xue, S.-S.  2004a, 35th COSPAR
Scientific Assembly, held 18-25 July 2004, Paris, France, 1064

\bibitem[Ruffini et al.(2004e)]{RBCFGX2} Ruffini, R., Bianco, C. L.,
Chardonnet, P., Fraschetti, F., Gurzadyan, V., \& Xue, S.-S.  2004b, Int. J.
Mod. Phys. D13, 843

\bibitem[Ruffini et al.(2003c)]{RBCFVX} Ruffini, R., Bianco, C. L., Chardonnet,
P., Fraschetti, F., Vitagliano, L., \& Xue, S.-S.  2003, astro-ph/0302557

\bibitem[Ruffini et al.(2002b)]{RBCFX1} Ruffini, R., Bianco, C. L., Chardonnet,
P., Fraschetti, F., \& Xue, S.-S.  2002, Ap.\ J.\  581, L19

\bibitem[Ruffini et al.(2003a)]{RBCFX2} Ruffini, R., Bianco, C. L., Chardonnet,
P., Fraschetti, F., \& Xue, S.-S.  2003, Int. J. Mod. Phys. D12, 173

\bibitem[Ruffini et al.(2004a)]{RBCFX3} Ruffini, R., Bianco, C. L., Chardonnet,
P., Fraschetti, F., \& Xue, S.-S.  2004, Third Rome Workshop on Gamma-Ray
Bursts in the Afterglow Era ASP Conference Series, Volume 312, Proceedings of
the conference held 17-20 September 2002, in Rome, Italy, eds. M. Feroci, F.
Frontera, N. Masetti, \& L. Piro, San Francisco: Astronomical Society of the
Pacific, 349

\bibitem[Ruffini et al.(2001a)]{RBFCX} Ruffini, R., Bianco, C. L., Fraschetti,
F., Chardonnet, P., \& Xue, S.-S.  2001, Nuovo Cim. B116, 99 (2001)

\bibitem[Ruffini et al.(2001b)]{RBFXC} Ruffini, R., Bianco, C. L., Fraschetti,
F., Xue, S.-S., \& Chardonnet, P.  2001, Ap.\ J.\  555, L107 \& L113 \& L117

\bibitem[Ruffini et al.(2004d)]{RFVX1} Ruffini, R., Fraschetti, F., Vitagliano,
L., \& Xue, S.-S.  2004, 35th COSPAR Scientific Assembly, held 18-25 July 2004,
Paris, France, 1316

\bibitem[Ruffini et al.(2005a)]{RFVX2} Ruffini, R., Fraschetti, F., Vitagliano,
L., \& Xue, S.-S.  2005, Int. J. Mod. Phys. D14, 131

\bibitem[Ruffini et al.(1998)]{RSWX1} Ruffini, R., Salmonson, J., Wilson, J.,
\& Xue, S.-S.  1998, Abstracts of the 19th Texas Symposium on Relativistic
Astrophysics and Cosmology, held in Paris, France, Dec. 14-18, 1998, eds. J.
Paul, T. Montmerle, \& E. Aubourg, Saclay: CEA

\bibitem[Ruffini et al.(1999a)]{RSWX2} Ruffini, R., Salmonson, J., Wilson, J.,
\& Xue, S.-S.  1999a, Ap.\ J.\ Suppl.\ 138, 511 

\bibitem[Ruffini et al.(1999b)]{RSWX3} Ruffini, R., Salmonson, J., Wilson, J.,
\& Xue, S.-S.  1999b, Astron.\ Astrophys.\ 350, 334

\bibitem[Ruffini et al.(2000)]{RSWX4} Ruffini, R., Salmonson, J., Wilson, J.,
\& Xue, S.-S.  2000 Astron.\ Astrophys.\ 359, 855

\bibitem[Ruffini \& Vitagliano(2002)]{RV} Ruffini, R., \& Vitagliano, L.  2002,
Phys. Lett. B545, 233

\bibitem[Ruffini, Vitagliano, \& Xue(2003a)]{RVX1} Ruffini, R., Vitagliano,
L., \& Xue, S.-S.  2003a, Phys. Lett. B559, 12

\bibitem[Ruffini, Vitagliano, \& Xue(2003b)]{RVX2} Ruffini, R., Vitagliano, L.,
\& Xue, S.-S.  2003b, Quantum Aspects of Beam Physics. Proceedings, Joint 28th
ICFA Advanced Beam Dynamics and Advanced and Novel Accelerator Workshop on
Quantum Aspects of Beam Physics (QABP03), Higashi Hiroshima, Japan, 7-11 Jan
2003, eds. P. Chen and K. Reil, Hackensack, USA: World Scientific, 295

\bibitem[Ruffini, Vitagliano, \& Xue(2003c)]{RVX3} Ruffini, R., Vitagliano, L.,
\& Xue, S.-S.  2003c, Quantum Aspects of Beam Physics. Proceedings, Joint 28th
ICFA Advanced Beam Dynamics and Advanced and Novel Accelerator Workshop on
Quantum Aspects of Beam Physics (QABP03), Higashi Hiroshima, Japan, 7-11 Jan
2003, eds. P. Chen and K. Reil, Hackensack, USA: World Scientific, 303

\bibitem[Ruffini, Vitagliano, \& Xue(2003d)]{RVX4} Ruffini, R., Vitagliano,
L., \& Xue, S.-S.  2003d, Phys. Lett. B573, 33

\bibitem[Ruffini \& Xue(1998)]{RX} Ruffini, R., \& Xue, S.-S.  1998, Abstracts
of the 19th Texas Symposium on Relativistic Astrophysics and Cosmology, held in
Paris, France, Dec. 14-18, 1998, eds. J. Paul, T. Montmerle, \& E. Aubourg,
Saclay: CEA

\bibitem[Ruffini et al.(2002a)]{RXBFC} Ruffini, R., Xue, S.-S., Bianco, C. L.,
Fraschetti, F., \& Chardonnet, P.  2002, La Recherche 353, 30

\bibitem[Sauter(1931)]{Sch1} Sauter, F.  1931, Z. Phys. 69, 742

\bibitem[Schwinger(1951)]{Sch4} Schwinger, J. S.  1951, Phys. Rev. 82, 664

\bibitem[Torres \& Anchordoqui(2004)]{TA} Torres, D. F., \& Anchordoqui, L. A. 
2004, Rep. Prog. Phys. 67, 1663

\bibitem[Usov(1998)]{Usov98} Usov, V. V.  1998 Phys.\ Rev.\ Lett.\  80, 230

\bibitem[Usov(2004)]{Usov04} Usov, V. V.  2004 Phys. Rev.\ D 70, 067301

\bibitem[Usov, Harko, \& Cheng(2005)]{Usov05} Usov, V. V., Harko, T., \& Cheng,
K. S.  2006 Ap.\ J.\  620, 915

\bibitem[Vagenas(2006)]{Vag} Vagenas, E. C.  2006, gr-qc/0602107

\bibitem[Weisskopf(1936)]{Sch3} Weisskopf, V.  1936, K. Dan. Vidensk.
Selsk. Mat. Fys. Medd. 14, 6

\bibitem[Xue et al.(2005)]{XRFG} Xue, S.-S., Ruffini, R., Fraschetti, F., \&
Gurzadyan, V.  2005, American Astronomical Society Meeting 207, \#158.02

\end{thebibliography}
\end{document}